\documentclass[aps, prb, superscriptaddress, reprint, nobibnotes, twocolumn]{revtex4-2}
\usepackage{amsmath,amssymb}
\usepackage{color}
\usepackage{comment}
\usepackage{bm,graphicx,hyperref}
\hypersetup{%
  breaklinks = {true},
  citecolor = {blue},
  colorlinks = {true},
  linkcolor = {red},}

\usepackage{lineno}

\usepackage{cancel}

\begin{document}

\title{Dissipation and diffusion in one-dimensional solids}

\author{Harshitra Mahalingam}
\affiliation{Institute for Functional Intelligent Materials, National University of Singapore, 4 Science Drive 2, 117544, Singapore}

\author{B. A. Olsen}
\thanks{Corresponding author}
\affiliation{Yale-NUS College, 16 College Avenue West, 138527, Singapore}
\affiliation{Department of Physics, National University of Singapore, 2 Science Drive 3, 117551, Singapore}

\author{A. Rodin}
\thanks{Corresponding author}
\affiliation{Yale-NUS College, 16 College Avenue West, 138527, Singapore}
\affiliation{Centre for Advanced 2D Materials, National University of Singapore, 117546, Singapore}
\affiliation{Materials Science and Engineering, National University of Singapore, 117575, Singapore}

\begin{abstract}

Using a nonperturbative classical model for ionic motion through one-dimensional (1D) solids, we explore how thermal lattice vibrations affect ionic transport properties.
Based on analytic and numerical calculations, we find that the mean dissipation experienced by the mobile ion is similar to that of the non-thermal case, with thermal motion only contributing stochastic noise.
A nonmonotonic dependence of drag on speed, predicted in earlier work, persists in the presence of thermal motion.
The inverse relation between drag and speed at high speeds results in non-Fickian diffusion dominated by L\'{e}vy flights.
This suppression of drag at high speeds, combined with enhanced activation frequency, improves the particle mobility at high temperatures, where typical particles move faster.

\end{abstract}	

\maketitle

\section{Introduction}
\label{sec:Introduction}

Understanding ionic motion through solids~\cite{MahanRoth, Mehrer} has direct practical applications, such as in solid electrolytes for advanced batteries~\cite{Bachman2016, Manthiram2017, Famprikis2019}.
One major question pertaining to these solid electrolytes is: what makes a good ionic conductor?

\begin{figure}
    \centering
    \includegraphics[width = \columnwidth]{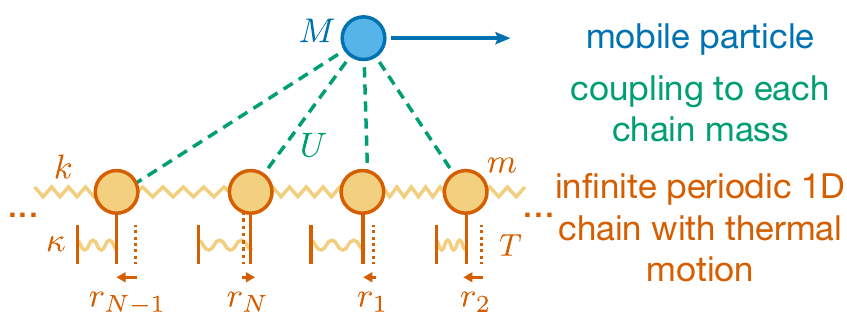}
    \caption{\emph{Schematic of the system.} A mobile particle of mass $M$ moves along a periodic 1D chain of $N\rightarrow \infty$ identical masses $m$, separated by distance $a$ at equilibrium.
    Each chain mass undergoes harmonic motion with spring constant $\kappa$ and displacement $r_i$, and couples to its neighbor with spring constant $k$.
    The $N$ oscillatory modes of the chain act as a bath, absorbing and imparting energy to the mobile particle due to the interaction $U$.
    The chain temperature $T$ dictates the homogeneous motion in $r_i$.}
    \label{fig:Schematic}
\end{figure}

We previously considered a simple model of ionic conductors~\citep{Mahalingam2023}: an infinite one-dimensional (1D) chain of masses interacting with a single mobile particle moving along the chain (see Fig.~\ref{fig:Schematic}).
Using classical equations of motion, we found that the particle experiences unconventional drag---it is non-linear and decreases at higher speeds.
This unconventional drag leads to multiple stable drift velocities when the system is subjected to a bias.

In this study, we introduce thermal motion to the chain, and use a combination of analytical and numerical methods to explore the role that temperature plays in ionic motion.
Thermal effects can drastically change the character of ionic motion, as shown by  molecular dynamics (MD) simulations~\cite{Deng2015, Deng2017, Carvalho2022}, which use \emph{ab initio} computations to predict the full dynamics of ionic conductors (but for very short evolution times).
At low temperatures, the conducting ions ``hop'' between energy minima in a sublattice, while at high temperatures, the sublattice ``melts" and leads to ``superionic flow"~\citep{Deng2017}.
Our complementary approach not only predicts long-time behavior of ionic conductors, but also permits an explicit connection between lattice-ion interaction properties and macroscopic transport behavior.

%

We show here that the mean dissipation of 1D ionic conductors in the presence of thermal motion is the same, to leading order, as the non-thermal case~\citep{Mahalingam2023}, with a correction that we predict analytically and confirm numerically.
We also show that bias-supported drift velocities persist in the presence of thermal fluctuations at sufficiently low system temperatures.
Finally, we demonstrate that unconventional drag in these 1D conductors leads not to typical Brownian motion, but rather to a L\'{e}vy flight,~\citep{Chechkin2008} giving rise to non-Fickian diffusion.

The results of our work are directly applicable to one-dimensional~\cite{Cho2022} and quasi-one-dimensional~\cite{Furusawa1993, Furusawa2000, Mansson2014, Volgmann2017} systems hosting mobile ions.
In addition, by focusing on a system that is somewhat analytically tractable, we are able to evaluate the validity of a number of simplifying assumptions.
Thus, the intuition gained here will be vital when studying higher-dimensional systems for which a full simplification-free treatment might be prohibitively difficult.

In Sec.~\ref{sec:EExchange}, we discuss how the classical equations of motion analyzed in \cite{Mahalingam2023} are modified in the presence of thermal motion, and outline the general contributions of fluctuation and dissipation terms.
In Sec.~\ref{sec:Dissipation}, we analyze how a mobile particle dissipates energy through interactions with the chain.
We discuss some simplifying assumptions that allow us to analytically derive the statistical properties of energy loss each time the particle passes a chain mass, and compare that single-pass model to numerical integration of the equations of motion.
We also explore how the interplay of thermal motion, dissipation, and bias influences drift velocities.
Finally, considering particles starting at rest, we calculate the thermalization and diffusive properties of this system in  Sec.~\ref{sec:Thermalization_Diffusion}.

\section{Energy Exchange}
\label{sec:EExchange}
\subsection{Model}
\label{sec:Model}

\begin{figure*}[ht]
    \centering
    \includegraphics[width = \textwidth]{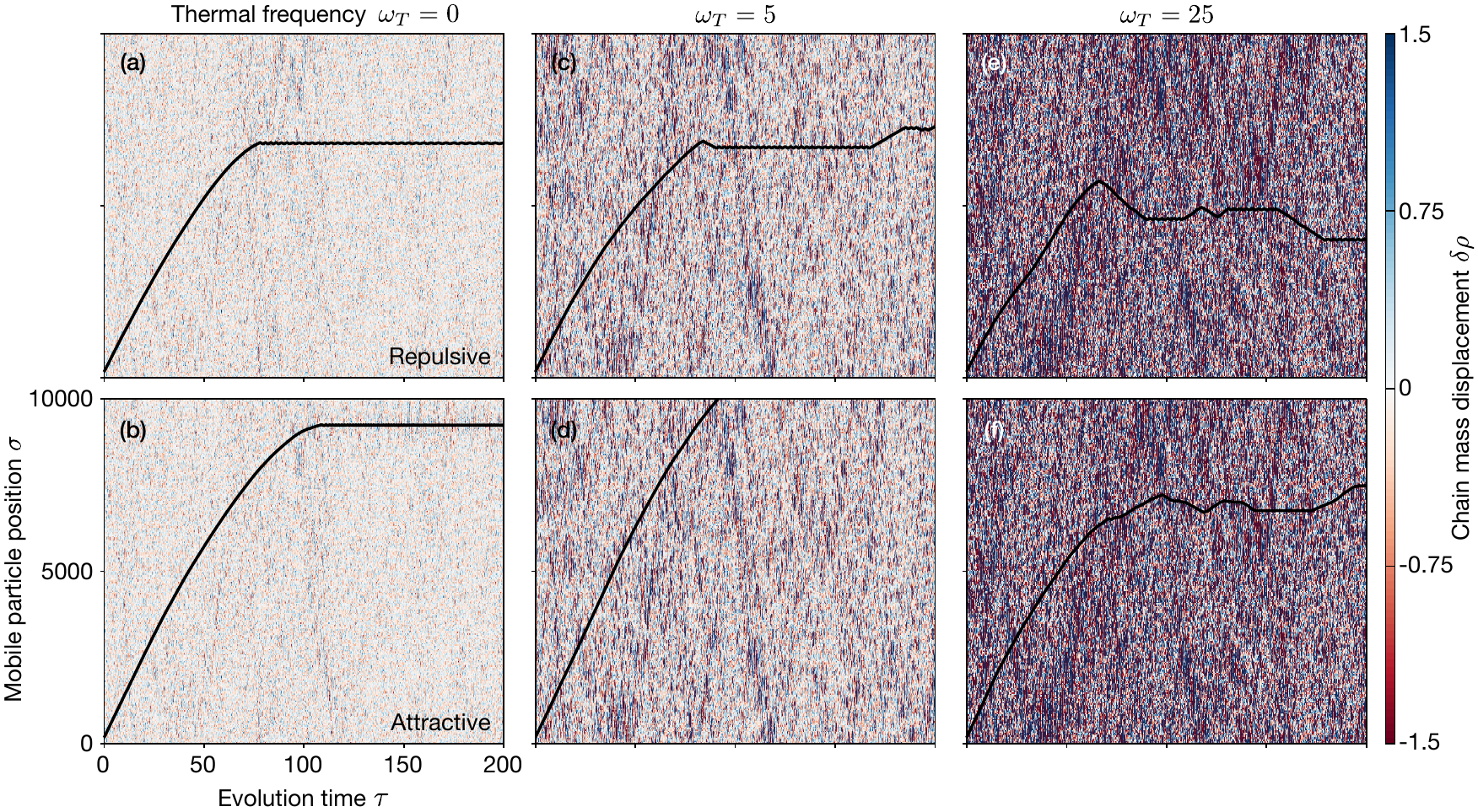}
    \caption{\emph{General example of fluctuation-dissipation.} Motion of a single particle for repulsive (top row) and attractive (bottom row) Gaussian chain-particle interaction across different temperatures (by column). All cases have chain spacing $\alpha = 40$, Gaussian width $\lambda = 4$, Gaussian amplitude $\Phi_0 = \pm 20$ and chain mass $\mu = 1$. The particles are initialized with particle speed $\dot{\sigma}_0 = 120$ at a point halfway between chain mass rest positions. The mobile particle trajectories $\sigma(\tau)$ are shown as black lines and exhibit fluctuation-dissipation behavior, where they lose energy to the chain and slow down, or gain energy from the chain and speed up. The displacement of 250 individual chain masses are shown with a heatmap, with the color scale saturated to highlight the chain displacement amplitudes across different temperatures.}
    \label{fig:General_Example}
\end{figure*}
To describe the motion of mobile particles through a framework of masses with vibrational modes, we follow the procedure used in Refs.~\citep{Rodin2022, Rodin2022a, Mahalingam2023}.
In this formalism, we derive equations of motion for the dimensionless chain mass displacements $\boldsymbol{\rho}$ and mobile particle positions $\boldsymbol{\sigma}$. All lengths are normalized by the quantum oscillator length associated with the slowest lattice mode, $l_\text{slow} = \sqrt{\hbar / m\Omega_\mathrm{slow}}$ with $\hbar$ the reduced Planck constant (e.g. the lattice spacing $\alpha=a/l_\text{slow}$). 
The dimensionless evolution time of the system, $\tau$, is normalized by $t_\text{slow}=2\pi/\Omega_\text{slow}$, the period of the slowest chain mode.
Similarly, all other dimensionless frequencies of the system are normalized by the slowest chain mode frequency: $\omega=\Omega/\Omega_\text{slow}$ (e.g. the fastest band frequency $\omega_\text{fast}=10$).
We define a thermal frequency $\omega_T=k_BT/\hbar\Omega_\text{slow}$, where $k_B$ is the Boltzmann constant and $\hbar$ is the reduced Planck constant. 
The normalized mobile particle mass is then $\mu = M /m$.
Using these normalized quantities, the dispersion of the chain's vibrational modes is given by
\begin{equation}
	\omega_j 
	= \sqrt{1 + \left(\omega_\mathrm{fast}^2-1 \right)\sin^2\left(\frac{\pi j}{N }\right)}\,,
	\label{eqn:Omega}
\end{equation}
and the trajectories of the framework masses and the mobile particles are related via the equations of motion

\begin{align}
\boldsymbol{\rho}(\tau)
   &= \sum_j\boldsymbol{\varepsilon}_j\sqrt{n_j + \frac{1}{2}}\sqrt{\frac{2}{\omega_j}}e^{-2\pi i \omega_j \tau + i\phi_j}
   \nonumber
   \\
   &-2\pi \int^\tau d\tau'
   \tensor{\Gamma}(\tau - \tau')
   \nabla_{\boldsymbol{\rho}}\Phi\left[\boldsymbol{\rho}(\tau'),\boldsymbol{\sigma}(\tau')\right]
   \,,
    \label{eqn:rho}
   \\
    \tensor{\Gamma}(\tau)
    &=\sum_j \boldsymbol{\varepsilon}_j \boldsymbol{\varepsilon}_j^\dagger\frac{\sin\left(2\pi\omega_j \tau\right)}{\omega_j}\,,
   \label{eqn:Gamma}
   \\
   \ddot{\sigma}_{j}(\tau) &= -(2\pi)^2\frac{1}{\mu}\sum_k\frac{d}{d\sigma_j}\Phi\left[\rho_k(\tau),\sigma_{j}(\tau)\right] \,,
   \label{eqn:sigma}
\end{align}
where the $j$ summation runs over the vibrational modes.
$\boldsymbol{\varepsilon}_j$ is the eigenvector of the $j$th vibrational mode with entries $\varepsilon_{g, j} = e^{2 \pi i g / N}/\sqrt{N}$ corresponding to the $g$th chain mass.
The first term in Eq.~\eqref{eqn:rho}, the homogeneous trajectory $\boldsymbol{\rho}_H$, describes the motion of the chain masses due to thermal fluctuations, given as a sum of normal modes $\boldsymbol{\rho}_H(\tau) = \sum_j \boldsymbol{\varepsilon}_j\zeta_j(\tau)$.
$\phi_j$ is the phase of the normal coordinate oscillation and $n_j$, a non-negative integer, is the excitation level of the $j$th mode following a Boltzmann distribution $e^{-n_j\omega_j / \omega_T}$.
Finally, $\Phi$ in Eqs.~\eqref{eqn:rho} and \eqref{eqn:sigma} is the interaction between the chain masses and the mobile particles. 

A key difference in our analysis compared to previous work is the interplay between thermal motion of the chain and the transport properties of the mobile masses. 
This stochastic behavior necessitates statistical analysis of many individual particle trajectories.

As it is composed of independent harmonics with random phases and thermally-distributed amplitudes, the homogeneous trajectory of the chain masses $\boldsymbol{\rho}_H(\tau)$ is a stationary Gaussian process.
Therefore, the statistical properties of $\boldsymbol{\rho}_H$ can be described using the means of $\rho_{H,g}$ and relevant covariances between the individual chain displacements.
Each mean $\langle\rho_{H,g}(\tau)\rangle $ vanishes due to the vanishing means of the normal coordinates.
The covariance matrix is given by $\tensor{C}(\tau-\tau') = \langle \boldsymbol{\rho}_H(\tau)\boldsymbol{\rho}_H^\dagger(\tau')\rangle = \sum_{jk}\langle \boldsymbol{\varepsilon}_j \boldsymbol{\varepsilon}^\dagger_k\zeta_j(\tau)\zeta_k^*(\tau')\rangle$, where the averaging is done over the mode phases $\phi_j$ and the energy levels $n_j$.
Because the normal modes are independent and their displacements average to zero, only $j = k$ contributes to the result, leading to

\begin{align}
   \tensor{C}(\tau)
 & =
  \sum_{j}
  \boldsymbol{\varepsilon}_j \boldsymbol{\varepsilon}_j^\dagger
  \langle \zeta_j(\tau)\zeta_j^*(0)\rangle
  \nonumber
  \\
  & =
  \sum_{j}
  \boldsymbol{\varepsilon}_j \boldsymbol{\varepsilon}_j^\dagger
   \frac{e^{-2\pi i \omega_j \tau}}{2\omega_j}
    \coth\left(\frac{\omega_j}{2\omega_T}\right)\,.
    \label{eqn:Correlation}
\end{align}
$\tensor{C}$ has a similar form to $\tensor{\Gamma}$ and is also a Toeplitz matrix so that its elements $C_{j,k}(\tau) = C_{j-k}(\tau)$.

As an instructive example, we computed the trajectory of a single mobile particle of mass $\mu = 1$ moving along a chain with $\omega_\mathrm{fast} = 10$ and $\alpha = 40$ for several different temperatures (see Fig.~\ref{fig:General_Example}).
We chose a Gaussian chain-particle interaction $\Phi(x) = \Phi_0 \exp(-x^2 / 2\lambda^2)$ with $\Phi_0 = \pm 20$ and $\lambda = 4$.
Here all the lengths are much greater than the quantum oscillator length of the slowest chain mode to ensure that the system can be seen as being firmly in the classical regime.
After initializing the mobile particle midway between two chain masses with initial speed $\dot{\sigma}_0 = 120$, we numerically integrated for $0\leq \tau \leq 200$.
The details of the computational procedure can be found in the Appendix of Ref.~\citep{Mahalingam2023}.
In Fig.~\ref{fig:General_Example}, the black lines show the particle's trajectory, while the heatmap displays the displacement $\delta \rho_j$ of each chain mass.
We see that the particle dissipates energy while also being subjected to the thermal fluctuation of the chain, which is more pronounced at higher temperatures.

For $\omega_T = 0$, shown in Fig.~\ref{fig:General_Example}(a) and (b) for repulsive and attractive chain-particle interaction, respectively, the dynamics are most similar to the non-thermal case.
The particle moves along the chain and gradually slows down, with greater deceleration in the repulsive case.
For repulsive interactions, the particle slows down in the vicinity of each chain mass, enhancing the energy transfer to the chain (for more detail, see Refs.~\citep{Rodin2022a, Mahalingam2023}).
Eventually, the particle becomes trapped in an energy minimum and stays there indefinitely.
The framework vibration at such a low temperature is not strong enough to kick the particle out of its trap or to greatly impact the particle's trajectory during the dissipative phase of the motion.

The impact of thermal fluctuations becomes more apparent in the case of $\omega_T = 5$---lattice vibrations can reactivate the particle after it becomes stuck, allowing it to continue moving along the chain (see Fig.~\ref{fig:General_Example}(c)).
This reactivation takes even less time at higher temperatures, as seen in Fig.~\ref{fig:General_Example}(e)-(f).
We also see that thermal motion can alter the rate of dissipation in the high-speed part of the trajectories.
Due to the stochastic nature of the thermal motion, we now turn our attention to the statistical properties of the particle-chain energy exchange.

\subsection{Energy fluctuation-dissipation}
\label{sec:Energy_fluctuation_dissipation}

To describe how particles lose energy as they move along the chain, we start by writing down the work done on the chain by the particles

\begin{equation}
\Delta(\tau) = 
\int^{\tau}  d\tau' \left\{-\nabla_{\boldsymbol{\rho}}\Phi\left[\boldsymbol{\rho}(\tau'),\boldsymbol{\sigma}(\tau')\right]\right\}^T\dot{\boldsymbol{\rho}}(\tau')\,,
\label{eqn:Delta}
\end{equation}
where the term inside the curly braces is the force on the chain masses and $\dot{\boldsymbol{\rho}}(\tau')$ their speed.
Splitting $\dot{\boldsymbol{\rho}}$ into its homogeneous and interaction-induced components (see Eq.~\eqref{eqn:rho}), we have

\begin{widetext}
\begin{align}
    \Delta(\tau) 
    =
       &  -
       \int^{\tau}  d\tau' 
   \nabla_{\boldsymbol{\rho}}\Phi\left[\boldsymbol{\rho}(\tau'),\boldsymbol{\sigma}(\tau')\right]^T\dot{\boldsymbol{\rho}}_H(\tau')
   +  \int^{\tau}  d\tau' 
  \nabla_{\boldsymbol{\rho}}\Phi\left[\boldsymbol{\rho}(\tau'),\boldsymbol{\sigma}(\tau')\right]^T
 \frac{d}{d\tau'} 2\pi\int^{\tau'}  d\tau''
  \Gamma(\tau' - \tau'')\nabla_{\boldsymbol{\rho}}\Phi\left[\boldsymbol{\rho}(\tau''),\boldsymbol{\sigma}(\tau'')\right] 
  \nonumber
  \\
   =
       &  -
       \int^{\tau}  d\tau' 
   \nabla_{\boldsymbol{\rho}}\Phi\left[\boldsymbol{\rho}(\tau'),\boldsymbol{\sigma}(\tau')\right]^T\dot{\boldsymbol{\rho}}_H(\tau')
   \nonumber
    \\
  &\quad+ 2\pi^2  \sum_j  \int^{\tau}  d\tau' \int^{\tau}  d\tau''
  \nabla_{\boldsymbol{\rho}}\Phi\left[\boldsymbol{\rho}(\tau'),\boldsymbol{\sigma}(\tau')\right]^T \boldsymbol{\varepsilon}_j 
\cos\left[2\pi\omega_j (\tau'-\tau'')\right]
\boldsymbol{\varepsilon}_j^\dagger \nabla_{\boldsymbol{\rho}}\Phi\left[\boldsymbol{\rho}(\tau''),\boldsymbol{\sigma}(\tau'')\right] \,.
    \label{eqn:Delta_Expanded}
\end{align}
\end{widetext}
When taking the derivative with respect to $\tau'$ in the second term, we used the fact that $\Gamma(0) = 0$ so that the only contribution comes from differentiating the integrand, which we did by explicitly writing out $\Gamma(\tau'-\tau'')$ following Eq.~\eqref{eqn:Gamma}.
We also exploited the symmetry of the $\tau'$ and $\tau''$ integrals in the second term after differentiation to set the upper integration limits to $\tau$ for $\tau'$ and $\tau''$ integrals, while also adding a factor of $1/2$.
Rewriting the $\cos(x) = (e^{ix} + e^{-ix}) / 2$ part of the last term, we have

\begin{align}
    \Delta(\tau) 
   =
       &  -
       \int^{\tau}  d\tau' 
   \nabla_{\boldsymbol{\rho}}\Phi\left[\boldsymbol{\rho}(\tau'),\boldsymbol{\sigma}(\tau')\right]^T\dot{\boldsymbol{\rho}}_H(\tau')
   \nonumber
    \\
  +& 2\pi^2  \sum_j  \left|\int^{\tau}  d\tau' 
  \nabla_{\boldsymbol{\rho}}\Phi\left[\boldsymbol{\rho}(\tau'),\boldsymbol{\sigma}(\tau')\right]^T \boldsymbol{\varepsilon}_j 
  e^{2\pi i\omega_j \tau'}\right|^2 \,.
    \label{eqn:Delta_Fluctuation_Dissipation}
\end{align}
This form of $\Delta(\tau)$ clearly indicates the fact that the second term is always non-negative, corresponding to energy transfer from the particles to the chain.
In contrast, the first term can be positive or negative, allowing us to write $\Delta =\Delta_\mathrm{fluc}+\Delta_\mathrm{diss}$ as a sum of ``fluctuation'' and ``dissipation'' components.
Averaging the fluctuation component over the thermal motion yields zero, which can be seen by writing the velocity vector as a combination of normal modes (as in Eq.~\eqref{eqn:rho}):

\begin{widetext}
\begin{align}
    \langle \Delta_\mathrm{fluc}(\tau)\rangle &= 
    \left\langle
       \int^{\tau}  d\tau' 
   \nabla_{\boldsymbol{\rho}}\Phi\left[\boldsymbol{\rho}(\tau'),\boldsymbol{\sigma}(\tau')\right]^T\sum_j\boldsymbol{\varepsilon}_j\sqrt{n_j + \frac{1}{2}}\sqrt{\frac{2}{\omega_j}}e^{-2\pi i \omega_j \tau' + i\phi_j}
   2\pi i \omega_j
   \right\rangle
   \nonumber
   \\
   &= 
       \int^{\tau}  d\tau' 
   \sum_j
   2\pi i \omega_j
   \sqrt{\frac{2}{\omega_j}}
   e^{-2\pi i \omega_j \tau' }
    \left\langle
    \sqrt{n_j + \frac{1}{2}}
    e^{i\phi_j}
\nabla_{\boldsymbol{\rho}}\Phi\left[\boldsymbol{\rho}(\tau'),\boldsymbol{\sigma}(\tau')\right]^T
   \right\rangle
   \boldsymbol{\varepsilon}_j = 0\,.
   \label{eqn:Fluctuation_Mean}
\end{align}
\end{widetext}

Here, the contribution to $\boldsymbol{\rho}(\tau')$ inside $\Phi$ from a single harmonic goes as $1 / \sqrt{N}$, so the phase-dependence of the force term can be neglected for every mode.
Consequently, averaging $e^{i\phi_j}$ over $\phi_j$ gives 0. 

We have shown that as the particles move, on average, they lose energy due to their interaction with the chain.
However, because of the chain's thermal motion, the actual value of $\Delta(\tau)$ is probabilistic, so we also compute the variance to see how it compares to the mean.
Squaring Eq.~\eqref{eqn:Delta_Fluctuation_Dissipation} produces a cross term, which vanishes for the same reason as Eq.~\eqref{eqn:Fluctuation_Mean}, leading to
\begin{equation}
    \mathrm{Var}\left[\Delta(\tau)\right]
    =
    \langle \Delta_\mathrm{fluc}^2(\tau)\rangle
    +
    \langle \Delta_\mathrm{diss}^2(\tau)\rangle
    -
    \langle \Delta_\mathrm{diss}(\tau)\rangle^2\,.
    \label{eqn:Var_Delta}
\end{equation}
In the absence of thermal motion, the first term vanishes and the last two terms become identical, leading to zero variance, so $\Delta$ is deterministic.

\section{Dissipation}
\label{sec:Dissipation}
\subsection{Single-pass dissipation}
\label{subsec:Single_pass_dissipation}

Despite the physical intuition offered by the terms in Eq.~\eqref{eqn:Delta_Fluctuation_Dissipation}, it is not solvable without already knowing the trajectories $\boldsymbol{\sigma}(\tau)$ and $\boldsymbol{\rho}(\tau)$.
To gain insight into the properties of $\Delta$ without resorting to full numerical integration, we will make a few key assumptions.

In our previous work, we considered a scenario without thermal motion ($\boldsymbol{\rho}_H = 0$)~\citep{Mahalingam2023}.
Here, we take the opposite limit and assume that the motion of the chain masses is dominated by the thermal component.
Consequently, we replace $\boldsymbol{\rho}$ by $\boldsymbol{\rho}_H$ inside the interaction terms in the expressions of $\Delta(\tau)$.
With this assumption, we expect the magnitude of $\Delta_\mathrm{fluc}$ to be much larger than $\Delta_\mathrm{diss}$.

Focusing on the interaction between a single mobile particle and a single chain mass, we then write the interaction as $\Phi(\rho_H  - \sigma)$.
Integrating the first term of Eq.~\eqref{eqn:Delta_Fluctuation_Dissipation} by parts and setting the interaction to zero at $\tau = \pm \infty$, we get

\begin{align}
    \Delta
      =
       &  
       \int  d\tau \,
        \dot{\sigma}(\tau)\Phi'[\rho_{H}(\tau)-\sigma(\tau)]
   \nonumber
    \\
  +&   \frac{2\pi^2}{N}\sum_j \int  d\tau   \int  d\tau'  
  e^{2\pi i\omega_j(\tau'-\tau'')}
  \nonumber
  \\
  \times &
  \Phi'\left[
  \rho_{H}(\tau)-\sigma(\tau)\right]
   \Phi'\left[
  \rho_{H}(\tau')-\sigma(\tau')\right]\,,
    \label{eqn:Delta_Single}
\end{align}
where the integration range has been set to $(-\infty, \infty)$.

Our goal now is to determine the statistical properties of $\Delta$ due to the probabilistic nature of the homogeneous motion.
The challenge lies in the fact that the particle trajectory $\sigma(\tau)$ also depends on the homogeneous motion.
Assuming the particle's speed remains constant during an interaction, $\sigma(\tau) \rightarrow \dot{\sigma}\tau$, simplifies the expressions considerably.
This assumption is reasonable if we consider the portion of the particle's trajectory far from trapping, where its kinetic energy is significantly larger than the potential variation created by its interaction with the chain.

Writing the interaction term as a Fourier transform $\Phi(x) = (N\alpha)^{-1}\sum_q \Phi_{q}e^{iqx}$ leads to a mean loss given by

\begin{align}
     \langle\Delta \rangle
  & = 
  \frac{1}{N\alpha}\int d\tau \dot{\sigma} \sum_q iq\Phi_q \left\langle e^{iq\rho_H(\tau)}\right\rangle e^{-iq\dot{\sigma}\tau}
  \nonumber
  \\
  &+\frac{2\pi^2}{N}\sum_j \int d\tau \int  d\tau'\,e^{2\pi i \omega_j(\tau-\tau')}
\sum_{qq'}\frac{-qq'}{N^2\alpha^2}\Phi_{q}\Phi_{ q'}
  \nonumber
  \\
  &\times
  \left\langle e^{iq\rho_{H}(\tau)}e^{iq'\rho_{H}(\tau')}
  \right\rangle
   e^{-iq\dot{\sigma}\tau}e^{-iq'\dot{\sigma}\tau'}\,.
   \label{eqn:Mean_Delta}
\end{align}

The multi-exponential expectation value can be written generally as $\left\langle \exp\left[\sum_l iq_l\rho_{H, g_l}(\tau_l)\right]\right\rangle$.
We define $\mathbf{q} = \begin{pmatrix} q_1, q_2,  q_3,  \dots \end{pmatrix}$, where each $q_l$ corresponds to an individual exponential inside the brackets (such as $q_1 = q$ in the first term of Eq.~\eqref{eqn:Mean_Delta}, and $q_1 = q$, $q_2 = q'$ in the second), and $\tilde{\boldsymbol{\rho}}_H=\begin{pmatrix} \rho_{H, g_1}(\tau_1), & \rho_{H, g_2}(\tau_2), & \dots \end{pmatrix}$ with $g_l$ corresponding to different chain masses.
In other words, we have an expectation value of exponentials involving homogeneous displacements of different chain masses at various times.
Using the covariance matrix in Eq.~\eqref{eqn:Correlation} we obtain

\begin{align}
    &\left\langle \exp\left[\sum_l iq_l\rho_{H, g_l}(\tau_l)\right]\right\rangle 
    \nonumber
    \\
    =& 
    \mathcal{N} \int \prod_l d\rho_{H, g_l}(\tau_l)
    \exp\left[i\mathbf{q}^T \tilde{\boldsymbol{\rho}}_H\right]
    \exp\left[-\frac{1}{2}\tilde{\boldsymbol{\rho}}_H^T\mathcal{C}^{-1}\tilde{\boldsymbol{\rho}}_H\right] 
    \nonumber
    \\
    =& \exp\left[-\frac{1}{2}\mathbf{q}^T\mathcal{C}\mathbf{q}\right]\,,
    \label{eqn:Exponential_Expectation}
\end{align}
where the matrix elements $\mathcal{C}_{j,k} = C_{g_j - g_k}(\tau_j - \tau_k)$ and $\mathcal{N}$ is the normalization constant for the multivariate Gaussian distribution.
In the case of a single pass, we are only interested in the matrix elements with $j = k$.

Using Eq.~\eqref{eqn:Exponential_Expectation}, we see that the average in the first term of Eq.~\eqref{eqn:Mean_Delta} is $\tau$-independent and Gaussian in $q$.
The $\tau$ integral in Eq.~\eqref{eqn:Mean_Delta} produces a Dirac delta function, setting $q\rightarrow 0$ and eliminating the fluctuating term as long as $\Phi_0$ is finite, in agreement with Sec.~\ref{sec:Energy_fluctuation_dissipation}.

Since the amplitude of the homogeneous motion of the chain masses is much larger than the deflection caused by their interaction with mobile particles, the variance of $\Delta$ will be dominated by the fluctuating term.
Using the constant-speed approximation, we write

\begin{align}
   \langle\Delta \rangle
  & =
  \langle\Delta_\mathrm{diss} \rangle
  = \frac{2\pi^2}{N}\sum_j\int  d\tau'' e^{2\pi i \omega_j \tau''}\mathcal{F}(\tau'') \,,
    \label{eqn:Delta_Mean_F}
\\
  \mathrm{Var}[\Delta]&\approx \langle\Delta^2_\mathrm{fluc}\rangle
  = \dot{\sigma}^2 \int   d\tau'' \mathcal{F}(\tau'') \,,\text{~with}
    \label{eqn:Delta2_Mean_F}
    \\
    \mathcal{F}(\tau'') & = 
     \sum_{qq'}\frac{-qq'}{N^2\alpha^2}\Phi_{q}\Phi_{q'} \int d\tau 
     e^{-iq\dot{\sigma}\tau}e^{-iq'\dot{\sigma}(\tau - \tau'')}
     \nonumber
  \\
  &\times
  \left\langle 
  e^{iq\rho_{H}(\tau)}e^{iq'\rho_{H}(\tau-\tau'')}
  \right\rangle
       \,,
       \label{eqn:F}
\end{align}    
and we performed a change of variables $\tau - \tau' = \tau''$.
To calculate the expectation value in $\mathcal{F}(\tau'')$, we use Eq.~\eqref{eqn:Exponential_Expectation}
\begin{align}
    &\left\langle 
  e^{iq\rho_{H}(\tau)}e^{iq'\rho_{H}(\tau-\tau'')}
  \right\rangle
  \nonumber
  \\
  =&   \exp\left[-\frac{1}{2}\begin{pmatrix}
      q & q'
  \end{pmatrix}
  \begin{pmatrix}
      C_0(0)&C_{0}(\tau'')
      \\
   C_{0}(\tau'')&C_0(0)
  \end{pmatrix}
  \begin{pmatrix}
      q \\ q'
  \end{pmatrix}\right]\,,
\end{align}
yielding a quantity that is independent of $\tau$ and allowing us to take the integral over $\tau$ in Eq.~\eqref{eqn:F}:

\begin{align}
    \mathcal{F}(\tau'') & = 
     \sum_{qq'}\frac{-qq'}{N^2\alpha^2}\Phi_{q}\Phi_{q'} 
   e^{iq'\dot{\sigma}\tau''}
     2\pi\delta[\dot{\sigma}(q + q')]
     \nonumber
  \\
  &\times
  \left\langle 
  e^{iq\rho_{H}(\tau)}e^{iq'\rho_{H}(\tau-\tau'')}
  \right\rangle
  \nonumber
  \\
  & = 
    \frac{1}{\dot{\sigma}} \sum_{q}\frac{q^2}{N\alpha}\Phi_{q}^2
   e^{-iq\dot{\sigma}\tau''}
   e^{-q^2\left(C_0(0)-C_0(\tau'')\right)}
\,,
  \label{eqn:F_eval}
\end{align}
where we assumed that $\Phi(x)$ is an even function so that $\Phi_q = \Phi_{-q}$.

\subsection{Mean loss}
\label{sec:Mean_loss}

Suppressing thermal fluctuations eliminates the Gaussian in Eq.~\eqref{eqn:F_eval}, making it possible to take the $\tau$ integral, followed by the momentum integration to obtain

\begin{equation}
      \langle \Delta\rangle \rightarrow  \Delta 
    = \frac{1}{2N}\sum_j \left(\frac{4\pi^2\omega_j}{\dot{\sigma}^2}\Phi_{\frac{2\pi\omega_j}{\dot{\sigma}}}\right)^2\,,
   \label{eqn:Delta_no_fluctuations}
\end{equation}
a result previously derived in Ref.~\citep{Mahalingam2023}.
Equation~\eqref{eqn:Delta_no_fluctuations} indicates that for large $\dot{\sigma}$, where $\Phi_{2\pi\omega_j/\dot{\sigma}}$ approaches a constant value,  $\Delta\propto\dot{\sigma}^{-4}$.
Intuitively, the reason behind the decay is that, at higher speeds, the particle and the chain mass interact for a shorter time, leading to a reduced energy transfer.
\begin{figure}
    \centering
    \includegraphics[width = \columnwidth]{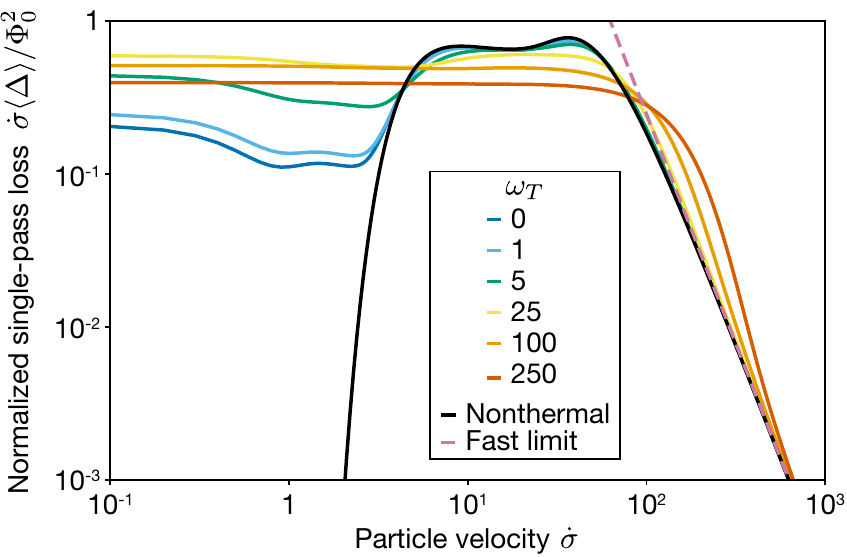}
    \caption{\emph{Mean energy loss over a large speed range.} Normalized mean single-pass loss $\dot{\sigma} \langle \Delta \rangle / \Phi_0^2$ as a function of speed for various temperatures (colored curves). At low speeds, $\langle \Delta \rangle \propto \dot{\sigma}^{-1}$, yielding nearly flat curves on the left side of the figure. The pink dashed line shows the $\dot{\sigma}^{-4}$ asymptote at high speeds.}
    \label{fig:Mean_loss}
\end{figure}
For $\dot{\sigma} \lesssim 2\pi\omega_j\Lambda$, where $\Lambda$ is the characteristic width of $\Phi(x)$, the suppressed Fourier term leads to a reduced energy absorption by the $j$th mode.
If the speed is sufficiently small so that $\dot{\sigma} \lesssim 2\pi\omega_j\Lambda$ for all the modes of the chain, the absorption is reduced drastically, leading to essentially no dissipation.
Conceptually, very slow particles do not deflect the chain mass so that the motion of the particle becomes essentially conservative.
We refer the reader to Ref.~\citep{Mahalingam2023} for a more in-depth discussion and numerical illustrations of Eq.~\eqref{eqn:Delta_no_fluctuations}.
In the absence of thermal fluctuations, wider potentials enhance dissipation for high particle speeds and suppress it for low $\dot{\sigma}$.

Reintroducing the thermal motion leads to several changes in the energy loss dependence on the particle speed.
In the case of a Gaussian interaction $\Phi(r) = \Phi_0\exp(-r^2 / 2\lambda^2)$, used in Fig.~\ref{fig:General_Example}, Eqs.~\eqref{eqn:Delta_Mean_F} and \eqref{eqn:Delta2_Mean_F} are given by
\begin{align}
    \langle \Delta \rangle 
   &= \frac{\pi^2 \lambda^2 \Phi_0^2 \sqrt{\pi}}{2\dot{\sigma} N} \sum_j \int d\tau \frac{e^{2 \pi i \omega_j \tau}}{(\lambda^2 + \tilde{C}(\tau))^{5/2}} \nonumber \\ 
   &\times \exp\left( -\frac{\dot{\sigma}^2 \tau^2 }{4(\lambda^2 + \tilde{C}(\tau))} \right)
   \left(2(\lambda^2 + \tilde{C}(\tau)) - \dot{\sigma}^2 \tau^2\right)    
   \label{eqn:gaussian_mean_delta}\, , \\
    \langle \Delta^2_\mathrm{fluc} \rangle &= \frac{\dot{\sigma} \Phi_0^2 \lambda^2 \sqrt{\pi}}{4} \int d\tau \frac{2(\lambda^2 + \tilde{C}(\tau)) - \dot{\sigma}^2 \tau^2}{(\lambda^2 + \tilde{C}(\tau))^{5/2}} \nonumber \\ 
    &\times \exp\left( -\frac{\dot{\sigma}^2 \tau^2 }{4(\lambda^2 + \tilde{C}(\tau))} \right) \, ,
    \label{eqn:gaussian_mean_var}
\end{align}
where $\tilde{C}(\tau) = C_0(0) - C_0(\tau)$.
We computed the mean numerically and present the results in Fig.~\ref{fig:Mean_loss}.

First, we observe that, at very high speeds, the temperature of the chain becomes irrelevant and all the curves collapse onto the $\Delta\propto \dot{\sigma}^{-4}$ asymptote.
This collapse is reasonable because at very high speeds, the chain mass does not shift much due to its thermal motion during the interaction window regardless of the temperature.

Next, we see that, at high speeds, the higher-temperature chains exhibit more dissipation.
Thermal motion effectively changes the Fourier transform of the interaction in Eq.~\eqref{eqn:F_eval} from $\Phi_q\rightarrow\tilde{\Phi}_q = \Phi_q e^{-q^2\frac{\tilde{C}(\tau)}{2}}$, which is narrower than $\Phi_q$ in Fourier space, and broader in real space.
Because the magnitude of $C_0(\tau)$ increases with temperature (see Appendix~\ref{sec:temperature_effect}), higher temperatures yield more broadening.
Since broader potentials enhance dissipation at high speeds, higher temperatures then lead to greater $\langle\Delta\rangle$.
\begin{figure}
    \centering
    \includegraphics[width = \columnwidth]{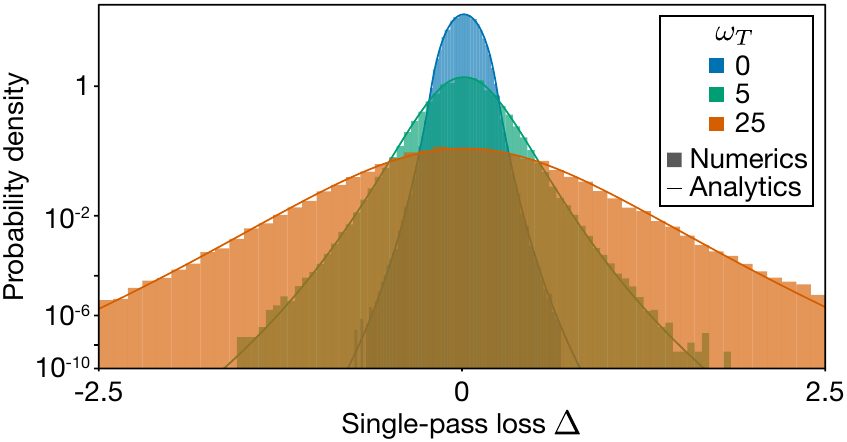}
    \caption{\emph{Numerical and analytical distribution of $\Delta$ over a single pass.} Distribution of single-pass energy loss $\Delta$ for 100,000 particles initialized at speed $\dot{\sigma} = 50$ for multiple temperatures. The probability densities of the numerical calculations are shown with shaded histograms, while the predicted probability densities from Eqs.~\eqref{eqn:gaussian_mean_delta} and \eqref{eqn:gaussian_mean_var} are shown with solid lines. The parameters here are $\Phi_0 = 2$, $\lambda = 4$, $\alpha = 40$ and $\omega_\mathrm{fast} = 10$. The number of bins is chosen to be constant across temperatures.}
    \label{fig:prob_density}
\end{figure}
\begin{figure}[t]
    \centering
    \includegraphics[width = \columnwidth]{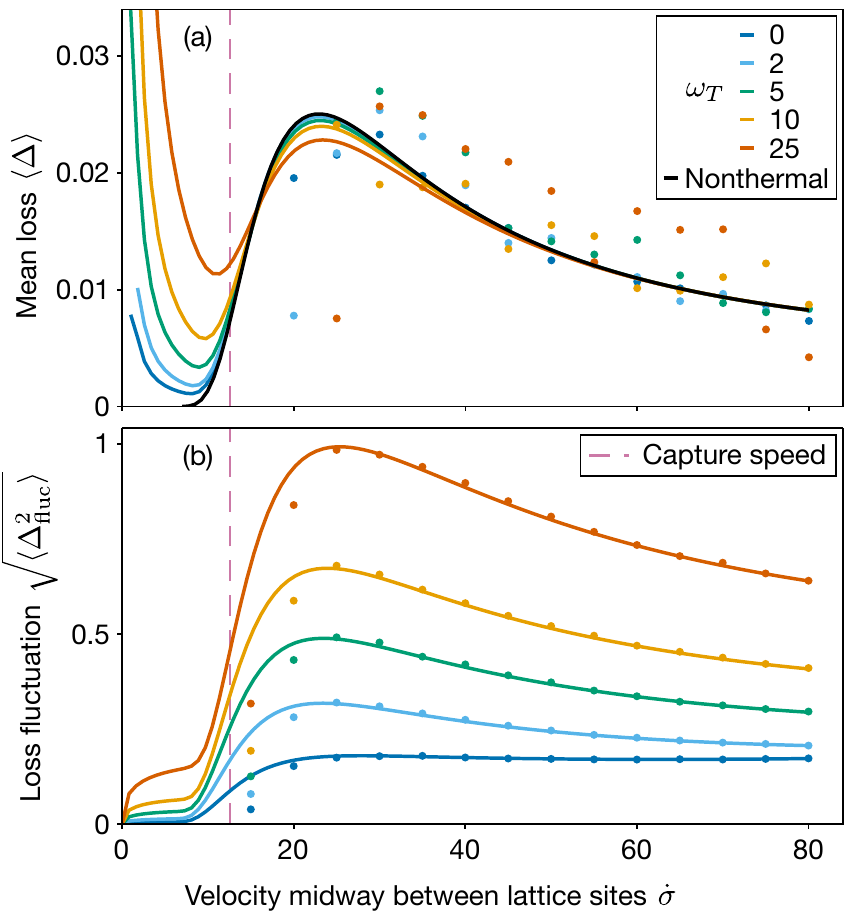}
    \caption{\emph{Mean and fluctuation of single-pass energy loss distributions.} Variation of (a) mean loss $\langle \Delta \rangle$ and (b) loss fluctuation $\sqrt{\langle \Delta_{\textrm{fluc}}^2 \rangle}$ during a single pass over a range of speeds and temperatures. The system parameters are $\Phi_0 = 2$, $\lambda = 4$, $\alpha = 40$ and $\omega_\mathrm{fast} = 10$. The analytical results are given by the solid curves while the numerical results are plotted as scatter points. The vertical dashed pink line is the capture speed of the system.}
    \label{fig:mean_var_Delta}
\end{figure}

At low speeds, thermal vibrations produce drastically different results from Eq.~\eqref{eqn:Delta_no_fluctuations}.
When $\dot{\sigma}\rightarrow 0$, we can drop the $e^{-iq\dot{\sigma}\tau''}$ term from Eq.~\eqref{eqn:F_eval} so that $\langle \Delta\rangle \propto \dot{\sigma}^{-1}$, as seen in Fig.~\ref{fig:Mean_loss}.
Although the dissipation diverges as $\dot{\sigma}^{-1}$ for all $\omega_T$'s , the prefactor is temperature-dependent and non-monotonic in $\omega_T$ (see Appendix~\ref{sec:temperature_effect} for more detail).

In the non-thermal case, dissipation vanishes at low speeds because the particle motion is essentially adiabatic.
At each time step, the quasi-stationary particle interacts with the chain,  modifying the chain modes.
However, due to the adiabatic theorem, the occupation of each mode remains unchanged.
Thus, when the particle passes the chain mass, the framework returns to its original configuration (with its original energy), leading to $\Delta = 0$.
With thermal motion, the process is no longer adiabatic because at each time step, the displacement between chain and mobile mass, and hence the interaction potential, varies.
Consequently, the occupation of the modes can change and the motion is no longer conservative, leading to a finite dissipation.
As $\omega_T$ increases, the system departs farther from adiabatic behavior, increasing $\Delta$.
\begin{figure*}
    \centering
    \includegraphics[width = \textwidth]{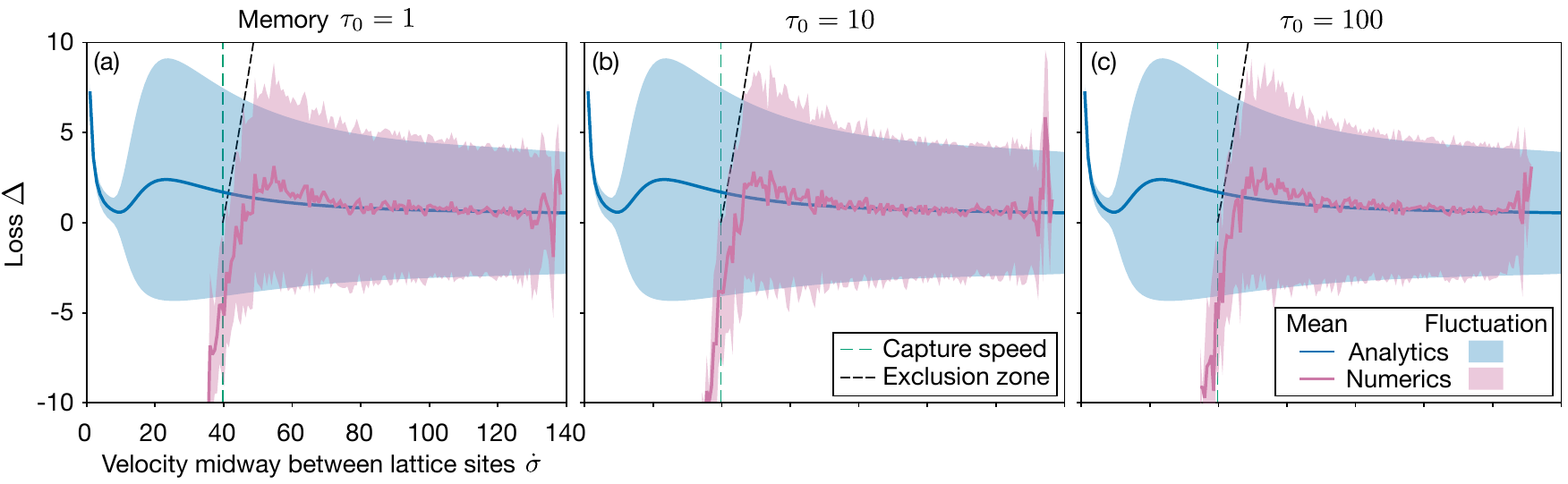}
    \caption{\emph{Energy losses $\Delta$ along full particle trajectories for different memories.} Distribution of energy losses after a single pass $\Delta$ for particles initialized at speed $\dot{\sigma} = 120$. The parameters for this system are the same as in Fig.~\ref{fig:General_Example}, with $\omega_\mathrm{fast} = 10$, $\Phi_0 = 20$, $\lambda = 4$, $\alpha = 40$ and temperature $\omega_T = 10$. From left to right, the three columns correspond to memory scales $\tau_0 = 1, 10$ and $100$. The blue curve and band give the analytical mean and fluctuation of the energy loss, respectively, based on Eqs.~\eqref{eqn:gaussian_mean_delta} and \eqref{eqn:gaussian_mean_var}, while the pink curve and band plot the numerical data from $\sim 250$ full trajectory runs. The  mean and variance of the numerical data is calculated by binning along the $\dot \sigma$ axis. The vertical dashed green line is the capture speed of the system. The dashed black curve plots the edge of the exclusion zone where a subset of particles are unable to pass the chain mass.}
    \label{fig:full_traj_loss}
\end{figure*}

At high temperatures, in agreement with Eq.~\eqref{eqn:F}, the potential profile becomes effectively broadened, leading to reduced dissipation at small speeds.
This suppression starts playing a role when the broadening becomes comparable to the characteristic width of the interaction.
Therefore, a competition between  potential broadening and adiabaticity breaking leads to a maximum in the dissipation for some intermediate $\omega_T$, as seen in Fig.~\ref{fig:Mean_loss}.

\subsection{Numerical results}
\label{sec:Numerical_results}
To numerically validate the assumptions made deriving Eqs.~\eqref{eqn:Delta_Mean_F}, \eqref{eqn:Delta2_Mean_F} and \eqref{eqn:F}, we chose a system with parameters $\omega_\mathrm{fast} = 10, \lambda =4, \Phi_0 = 2$ and $\alpha = 40$, and initialized mobile particles with speed $\dot{\sigma}_0 = 50$ midway between the rest positions of two chain masses. We then evolved the system forward in time for $\tau = 1.5 \times \alpha / \dot{\sigma}$, which is sufficiently long for the particle to pass a single chain mass and reach the next midpoint. The energy loss $\Delta$ over a single pass was calculated by taking the difference in kinetic energy $T = (\mu \dot{\sigma}^2) / (8 \pi^2)$ between the final and initial midpoints.
To average over homogeneous motion, we performed the same single-pass simulation over 100,000 realizations of thermal motion at a given temperature to obtain a distribution of $\Delta$ values.

Since the homogeneous motion of the chain masses $\boldsymbol{\rho}_H$ is a stationary Gaussian process, we expect the distribution of single-pass energy losses to be normally distributed around $\langle \Delta \rangle$ with variance $\langle \Delta^2_\textrm{fluc}\rangle$. 
Figure~\ref{fig:prob_density} shows numerical and predicted single-pass $\Delta$ probability density distributions for various temperatures. 
The numerical and predicted curves agree very well, confirming that the chain homogeneous motion is indeed a Gaussian process and validating the assumption $\boldsymbol{\rho} \approx \boldsymbol{\rho}_H$ made in Eq.~\eqref{eqn:Delta_Mean_F} and \eqref{eqn:Delta2_Mean_F}. As expected, the variance of the loss distribution increases with temperature. Compared to the variance, the value of the mean is small and does not appear to vary strongly with temperature. 

Having confirmed that the $\Delta$ distributions for individual speeds are normally distributed, we can repeat the same procedure across a larger range of speeds and temperatures. Figure~\ref{fig:mean_var_Delta} shows the mean loss $\langle \Delta(\dot{\sigma}) \rangle$ and loss fluctuation $\sqrt{\langle \Delta^2_{\textrm{fluc}}(\dot{\sigma}) \rangle}$ of single-pass energy loss distributions for various speeds and temperatures. 
The solid lines are analytic results and the points are numerical data. We see that the loss fluctuations agree very well with the analytical predictions, though this agreement becomes worse at particle speeds close to the capture speed. 
In this regime, the assumption that the particle travels at a constant speed breaks down due to interaction-induced velocity fluctuation (see Ref.~\cite{Mahalingam2023}).
While the values of $\langle \Delta \rangle$ obtained from numerical simulations do not agree closely with the analytic predictions, the large fluctuations suggest that it is unlikely for the mean values to have converged over 100,000 data points.
Additionally, since interactions induce velocity fluctuations, we expect a shift in the curve to the right (left) for repulsive (attractive) interactions---this phenomenon is discussed extensively in Ref.~\cite{Mahalingam2023}.

\subsection{Dissipation along full trajectory}
\label{sec:dissipation_full_traj}
We now turn to the dissipation experienced by a particle which passes multiple chain masses.
In the non-thermal case, $\Delta$ for a single pass differs significantly at slower speeds from the loss-per-pass observed in longer trajectories because of the cumulative effect of recoil~\cite{Mahalingam2023}.
In the previous subsection, we confirmed that, on the scale of a single pass, the chain motion is dominated by thermal vibrations and recoil is negligible when computing the force. 
Thus, the recoil term should play a much less significant role in modifying $\Delta$ during a multi-pass trajectory at finite temperature.
In other words, we expect the energy loss along a multi-pass trajectory to be governed on the scale of single passes and the dissipation statistics to remain the same as in the single-pass case.
Such a particle would essentially experience a random walk in speed, with the $\Delta$ distribution for the next pass only determined by the particle's current speed. 

To confirm the minimal role of the recoil term, we compared the energy loss distributions along full trajectories for different memory scales $\tau_0$ (so the limit of the integral in Eq.~\eqref{eqn:rho} becomes $\tau - \tau_0$). 
Initializing a mobile particle midway between chain mass rest positions at speed $\dot{\sigma} = 120$ within a periodic box, we evolved the system forward in time until the particle speed $\dot{\sigma} \leq 0$. 
We calculated the single-pass losses along the full trajectory by considering the kinetic energy of the particle at all midway points. 
This procedure was repeated for $\sim 250$ particles and the data binned by speed. 
In order to obtain a statistically significant sample size while keeping the computational cost low, we chose the same system parameters as in Fig.~\ref{fig:General_Example}---in particular, the larger amplitude $\Phi_0 = 20$ ensures a greater dissipation rate and allows for shorter simulation times.

Figure~\ref{fig:full_traj_loss}(a), (b) and (c) show the energy losses along full trajectories for memory scales $\tau_0 = 1$, 10 and 100, respectively. The pink curve and band show the mean and standard deviation of the binned numerical data, while the blue curve and band show the analytical predictions given by Eq.~\eqref{eqn:gaussian_mean_delta} and the square root of Eq.~\eqref{eqn:gaussian_mean_var}. 
At high speeds, both the mean loss and loss fluctuation of the numerical data agree well with the analytics.
At intermediate speeds, velocity variations during each pass lead to a shift of the mean loss and loss fluctuation. 

Closer to the capture speed, plotted as a vertical dashed line in Fig.~\ref{fig:full_traj_loss}, the numerical mean sharply decreases into negative values.
In this regime, we can only compute $\Delta$ for particles that are able to pass the chain mass, leading to a selection bias in the data. 
At low enough speeds, this exclusion zone significantly truncates the $\Delta$ normal distribution at its upper tail. 
In the case where the particle's total energy is small enough,  $\mu \dot{\sigma}^2 / (8 \pi^2) - \Delta < \Phi_0$, the particle is unable to complete the pass. 
The edge of the exclusion zone is then given by $\Delta_{\mathrm{ex}}(\dot{\sigma}) = \mu \dot{\sigma}^2 / (8 \pi^2) - \Phi_0$, plotted as the black dashed curve in Fig.~\ref{fig:full_traj_loss}. 
Using Eqs.~\eqref{eqn:gaussian_mean_delta} and \eqref{eqn:gaussian_mean_var} as the mean and fluctuation of a full normal distribution, we can then obtain the mean and fluctuation of the distribution truncated at the exclusion zone edge.
The details of the calculation are discussed in Appendix~\ref{sec:exclusion_zone}.
In short, we see excellent agreement between the analytic and numerical curves, confirming that the energy loss distribution close to the capture speeds becomes truncated due to trapping.
The exclusion leads to the numerical mean reflecting not the dissipative properties of the system, but rather the statistics of particle trapping---$\Delta$ now being a distribution leads to a capture speed that is also a distribution rather than a deterministic value.
\begin{figure*}
    \centering
    \includegraphics[width = \textwidth]{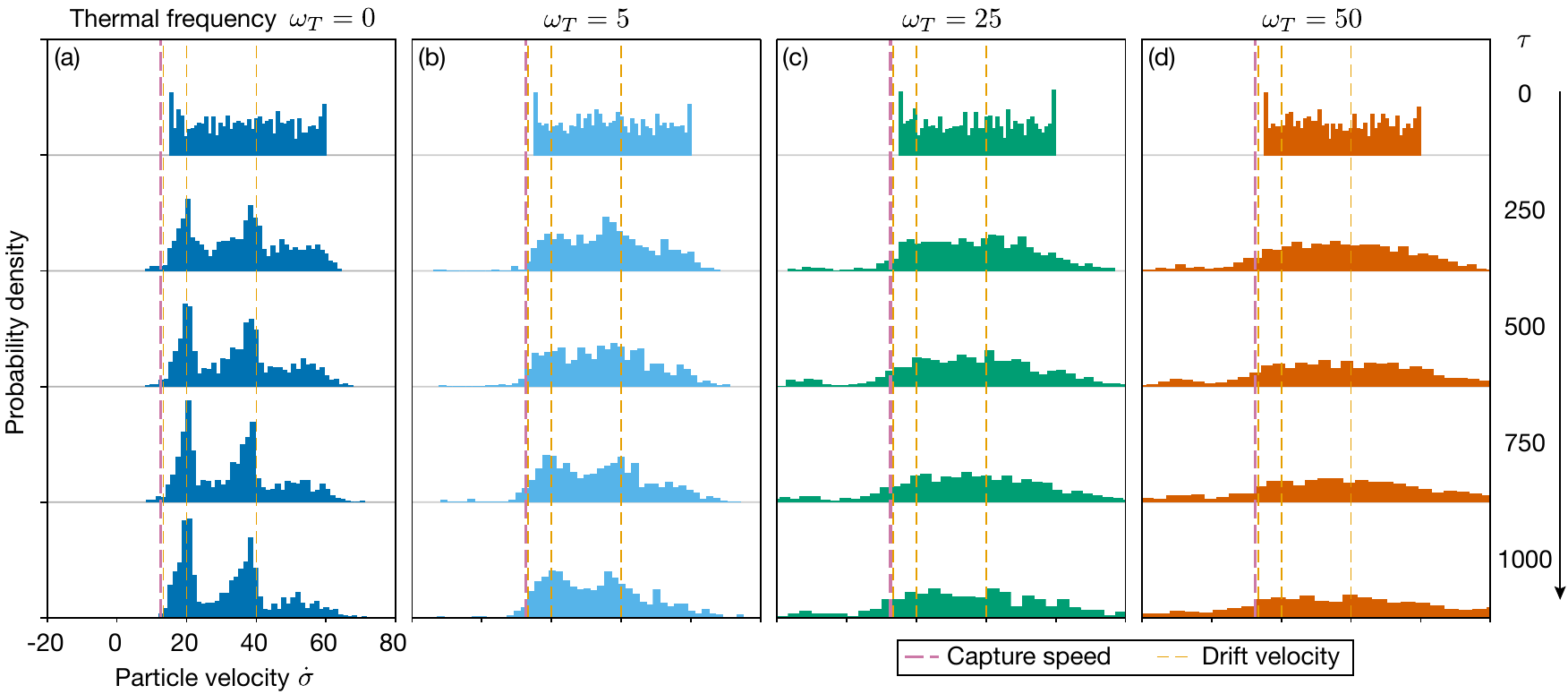}
    \caption{\emph{Bias-supported drift velocities.} Evolution of velocity distributions of 1000 particles after $\tau=250$ intervals for memory $\tau_0 = 10$. The parameters for these simulations are $\Phi_0 = 2$, $\lambda = 4$, $\alpha = 40$, $\omega_\mathrm{fast} = 10$, and bias $\beta = 0.01$. The particles were initialized midway between chain masses with speeds randomly sampled from an integer uniform distribution of $15 \leq \dot{\sigma} \leq 60$. From left to right, the columns correspond to temperatures $\omega_T =0$, $5$, $25$, and $50$. 
    The orange dashed vertical lines show the predicted drift velocities $\alpha / n$ which lie above the capture speed, shown as pink dashed vertical lines.}
    \label{fig:drift_velocity}
\end{figure*}

Comparing the different memory scales in Fig.~\ref{fig:full_traj_loss}(a), (b) and (c), we see that the memory term does not have a significant impact on the particle behavior, so we need only consider a memory scale that spans the time needed for a single pass. 
While individual trajectories with different memory scales are not identical, their statistics are the same, and are well-predicted by single-pass analytics. 

As a final check that the particle behavior is governed on the memory scale of a single pass, we constructed a random walk model in speed with a speed-dependent probability distribution. 
The details of the model are outlined in Appendix~\ref{sec:random_walk}.
The random walk offers us the ability to consider particle trajectories with and without the exclusion zone. 
In other words, when sampling $\Delta$ at the current step, we can choose to keep all values, or exclude values for which the particle has insufficient energy to pass the chain mass.
Without excluding any particles, the random walk model closely matches the analytic predictions in Fig.~\ref{fig:full_traj_loss}, while employing the exclusion zone causes the random walk to closely match the numerical calculations.
Since the random walk model replicates the statistical properties of particle behavior, it can be used to efficiently predict useful parameters such as stopping power.
\begin{figure*}
    \centering
    \includegraphics[width = \textwidth]{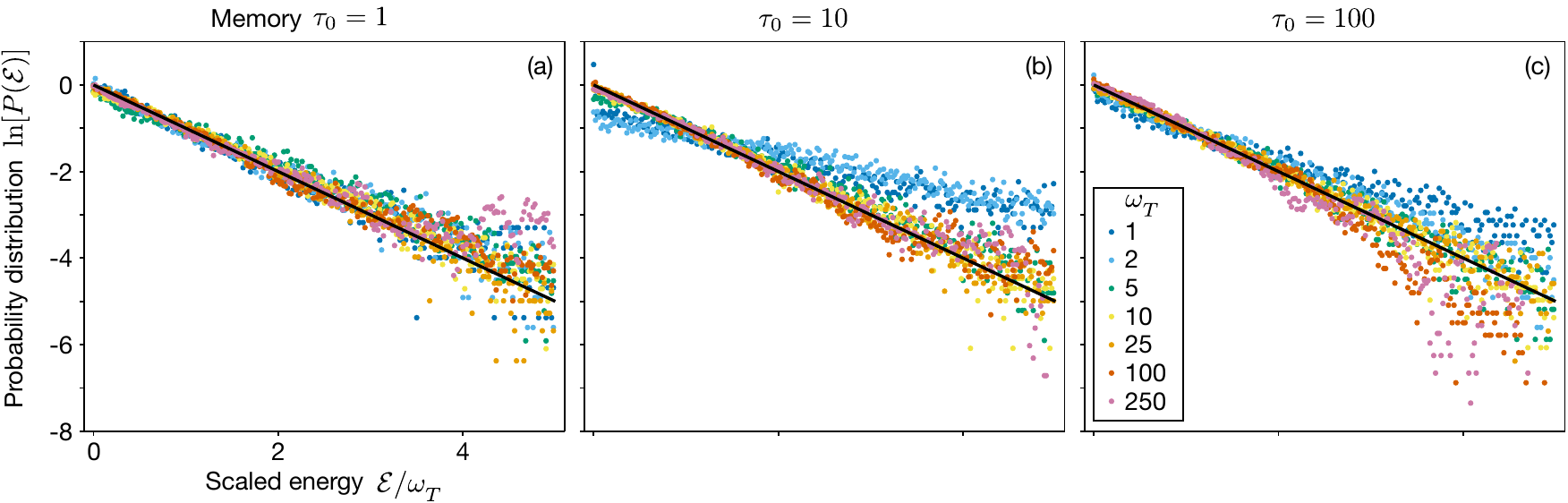}
    \caption{\emph{Energy probability distribution for different temperatures and memory scales.} Kinetic energy probability distribution per particle $\mathcal{E}$ vs. the kinetic energy normalized by effective chain temperature $\omega_T$. The system parameters are $\omega_{\mathrm{fast}} = 10$, $\mu = 1$, $\Phi_0 = 20$, $\lambda = 1$ and $\alpha = 10$. The particle energies are calculated midway between chain masses where the chain-particle interaction is negligible. From left to right, the three columns correspond to memory scales $\tau_0 = 1, 10$ and $100$. The black lines have slope $-1$, corresponding to Boltzmann distributions. For higher temperatures, the particles thermalize with the chain, even for short $\tau_0$.}
    \label{fig:Thermalization}
\end{figure*}

\subsection{Bias-aided drift velocities}
For a non-thermal system in the long-time limit, only certain chain modes absorb energy from the mobile particle~\cite{Mahalingam2023}.
This absorption is enhanced at particular speeds, leading to peaks in the mean dissipation per pass $\bar\Delta$ at speeds $\dot{\sigma} = \alpha / n$ where $\alpha$ is the lattice spacing and $n$ is a positive integer. 
In the presence of a bias $\beta$ (so the potential experienced by the mobile particle increases linearly by $\beta$ for a translation of $\alpha$), the particle experiences an energy loss of $\bar{\Delta}(\dot{\sigma}) - \beta$ over a pass---if this value is positive, the particle slows down and if this value is negative, the particle speeds up. 
If $\bar{\Delta}(\dot{\sigma}) - \beta = 0$, however, the particle maintains its speed.
These peaks in $\bar\Delta$ allow the mobile particle to experience stable drift velocities in the vicinity of $\dot{\sigma} = \alpha / n$, which we confirmed numerically. 

We have shown that the introduction of thermal motion leads to a statistical description of dissipation, with the expected energy loss value modulated by temperature. 
For speeds above the capture speed, the mean loss is very similar to that of the non-thermal case, as seen in Fig.~\ref{fig:mean_var_Delta}(a). 
Since averaging over the homogeneous motion results in a mean energy loss close to the non-thermal chain, the mean dissipation per pass $\bar{\Delta}$ for low temperatures remains similar in value to the non-thermal scenario. 
This suggests that for low enough temperatures, the drift velocities can survive in the thermal regime.

To confirm this prediction, we performed numerical simulations for systems under a constant bias $\beta = 0.01$, with memory $\tau_0 = 10$. 
We initialized 1000 particles in 40 batches with speeds randomly sampled from an integer uniform distribution and evolved the system in a periodic box with 250 chain masses for $\tau=1000$.
To add the bias, we introduced a constant force $\beta / \alpha$ to the particle's equations of motion.
We then extracted the particle velocity distributions at intervals of $\tau = 250$ and repeated the procedure for multiple temperatures. 
Figure~\ref{fig:drift_velocity} shows the evolution of particle velocity distributions over $\tau = 1000$ at various temperatures. 
Figure~\ref{fig:drift_velocity}(a), (b), (c), and (d) correspond to temperatures $\omega_T = 0$, $5$, $25$, and $50$, respectively. 
The top row shows the initial (essentially identical) particle velocity distributions. 
The expected drift velocities $\dot{\sigma} = \alpha / n$, where $n$ is an integer, are plotted as orange vertical dashed lines. 
Only drift velocities above the capture speed, plotted as pink dashed lines, are shown. 
For the case of $\omega_T = 0$, we see a clear evolution from a uniform velocity distribution to a distribution with strong peaks around $\dot{\sigma} = 40$ and $20$, corresponding to $n = 1$ and $2$, respectively. 
As the temperature increases, these peaks remain at roughly the same velocities, but broaden until they can no longer be clearly distinguished.

Thus, for sufficiently low temperatures, simple macroscopic transport measurements of 1D or quasi-1D materials can reveal microscopic parameters (in this case, the lattice spacing).

\section{Thermalization and Diffusion}
\label{sec:Thermalization_Diffusion}

\subsection{Thermalization}
\label{subsec:thermalization}

In addition to modifying the dissipative properties of the 1D chain, thermal vibrations can also induce fluctuation in the mobile particle's trajectory.
For instance, an initially motionless particle will interact with the chain masses and gain energy.
In a similar model, where a particle interacts with an infinite chain but is also confined to a harmonic trap, we demonstrated that the particle's energy distribution is dependent on the chain temperature $\omega_T$~\cite{Rodin2022a}.
In particular, the harmonically-trapped particle's energy follows a Boltzmann energy distribution $P(\mathcal{E}) \propto \exp(-\mathcal{E}/\omega_T)$, where $\mathcal{E}$ is the total energy of the particle scaled by a characteristic energy and $\omega_T$ is the scaled chain temperature.
To verify that this result also holds in the present case, we perform a set of numerical simulations and extract the particle's energy.

We computed 20 independent single-particle trajectories, initially at rest, in a confined box spanning $\sim 100$ chain masses for a range of temperatures.
The confined box allows us to run simulations for an extended period of time without the possibility of a particle moving past the end of the chain we are keeping track of.
We chose $\Phi_0 = 20$, $\alpha = 10$, $\lambda = 1$, $\omega_\mathrm{fast} = 10$ and $\mu = 1$ as the system parameters, which should allow the system to thermalize within a reasonable simulation time.
In particular, the combination of a large interaction amplitude $\Phi_0$ and small lattice spacing $\alpha$ leads to a higher rate of energy transfer from the chain to the particles.
We initialized the particles at randomized positions within the box and evolved the system for $\tau = 500$.

To explore the effects of memory, we performed the simulations for three different values of $\tau_0$ while using the same homogeneous motion of the chain masses $\boldsymbol{\rho}_H$.
After calculating the particle trajectories, we computed their kinetic energies midway between chain mass equilibrium positions.
At these points, the interaction between the particles and the chain is essentially zero so the particles' total energy is approximately equal to their kinetic energy.
For each memory-temperature pair, we used the sampled energies to build a normalized histogram, from which we extracted a probability distribution $P(\mathcal{E})$.
These distributions are plotted in Fig.~\ref{fig:Thermalization}(a), (b) and (c) as functions of $\mathcal{E}/\omega_{T}$ for memories $\tau_0 = 1$, $10$ and $100$, respectively.

Plotting $\ln[P(\mathcal{E})]$ vs. $\mathcal{E}/\omega_{T}$, the distribution should be a line with slope $-1$.
Indeed, we observe that the probability distributions collapse onto the black lines in Fig.~\ref{fig:Thermalization}, corresponding to the Boltzmann profile.
Memory does not have a significant effect on the kinetic energy distributions, especially for temperatures $\omega_T \geq 1$, the energy of the lowest chain mode.
At temperatures close to the band minimum, zero-point motion is no longer negligible and we expect the effective temperature to be larger than $\omega_T$.
This effect is most clearly understood at $\omega_T = 0$, where the chain masses move due to zero-point motion, in contrast to classical zero-temperature case where all motion ceases.
This motion results in a non-zero effective temperature.

The effects of the finite memory are most evident at low temperatures.
As we discussed in our earlier work~\citep{Rodin2022a}, memory truncation produces an artificial motion of the chain masses, referred to as ``countermotion" when the chain mass moves in the direction opposite to where it was pushed by a particle at an earlier time.
This unphysical countermotion is related to the recoil of the chain masses, and can modify the effective temperature of the particles.
However, since the recoil is dominated by thermal vibrations, this anomalous scaling is only evident for the lowest system temperatures, and changes nontrivially with memory.
This effect is most evident in Fig.~\ref{fig:Thermalization}(b).

\subsection{Diffusion}
\label{sec:diffusion}
\begin{figure}
    \centering
    \includegraphics[width = \columnwidth]{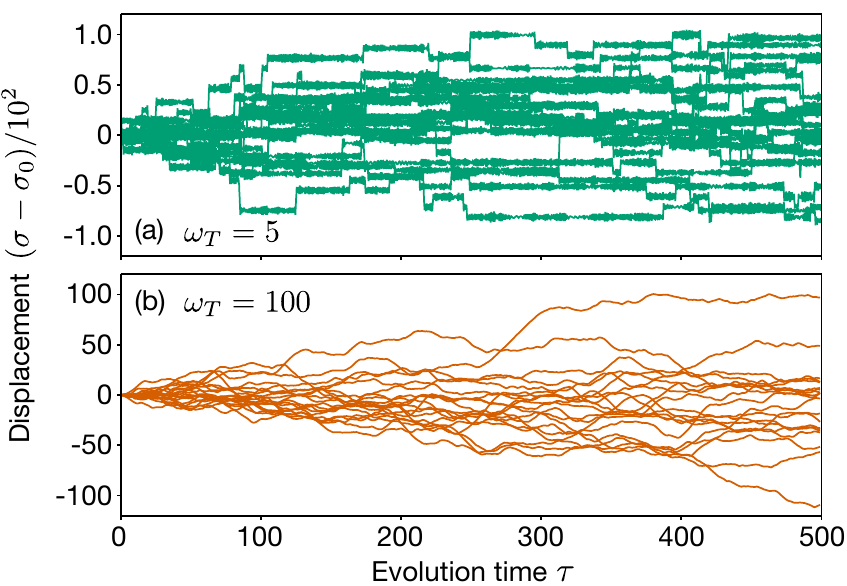}
    \caption{\emph{Displacement of 20 independent single-particle trajectories.} Displacements over time of 20 particles for system parameters $\omega_{\mathrm{fast}} = 10$, $\mu = 1$, $\Phi_0 = 20$, $\lambda = 1$ and $\alpha = 10$. The memory scale used is $\tau_0 = 100$. At low temperature $\omega_T=5$ (a), particles spend most of their time trapped between chain masses, and occasionally travel short distances. For high temperature $\omega_T=100$ (b), the particles spend very little time trapped, and their displacements appear smooth (though the overall scale for displacement is larger).}
    \label{fig:unfolded_traj}
\end{figure}

We now turn to the transport properties of particles initially at rest.
In the previous subsection, we found that truncating the memory for such particles can alter the energy distribution, so we restrict our analysis to long memory $\tau_0=100$.
We perform 20 independent single-particle simulations, in which we initialize the particle at rest randomly located within the boundaries of a confining box and evolve the system for $\tau = 500$ at several different chain temperatures.
In Figure~\ref{fig:unfolded_traj}, we show the subsequent displacements for systems at low and high temperatures ($\omega_T = 5$ and $100$, respectively).
For $\omega_T=5$, we see that the particles spend most of their time trapped between chain masses, and occasionally travel short distances.
In contrast, the particles at $\omega_T=100$ are rarely trapped, and their resulting trajectories look comparatively smooth.
As expected, the overall displacement for the higher temperature particles is larger.

The decreasing dissipation at high speeds, as seen in Figs.~\ref{fig:Mean_loss} and \ref{fig:mean_var_Delta}, means that typical particles at higher temperatures dissipate less.
For example, the average kinetic energy at $\omega_T = 250$ translates to $\dot{\sigma}\approx 140$, while $\omega_T = 100$ gives $\dot{\sigma}\approx 90$ and $\omega_T = 25$ yields $\dot{\sigma}\approx 44$.
All these speeds are on the right of the maximum in $\langle \Delta\rangle$ with the faster-moving particles experiencing less dissipation.
Hence, at higher temperatures, not only do particles become activated more frequently, but they also travel farther because of increased speed accompanied by reduced drag.
\begin{figure}
    \centering
    \includegraphics[width = \columnwidth]{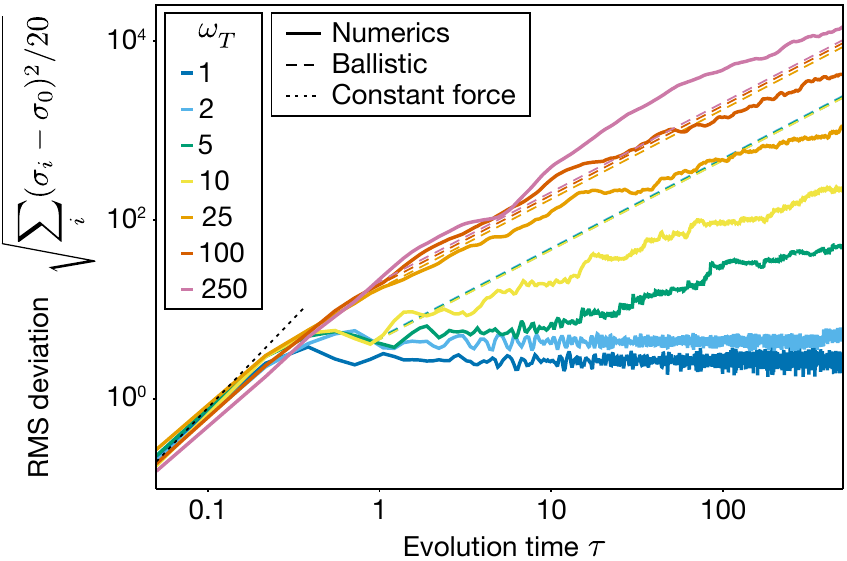}
    \caption{\emph{Root mean squared displacement for 20 independent single-particle trajectories.} The evolution of RMSD averaged over 20 particles initially at rest.
    The chain-particle interaction has Gaussian amplitude $\Phi_0 = 20$ and Gaussian width $\lambda = 1$, while the chain spacing is $\alpha = 10$.
    The black dotted line at short times has a slope of $+2$, while the dashed lines are given by $D_0 \tau$, where $D_0$ is a constant that is fit to the first value of each curve.
    These dashed lines of slope $+1$ correspond to ballistic transport.
    The constant force trend at small values of $\tau$ corresponds to the particles experiencing a slowly-changing potential profile.}
    \label{fig:diffusion_rmsd}
\end{figure}

Since this system exhibits anomalous dissipation, we expect a collection of particles to exhibit non-Fickian diffusion.
Using the calculated displacements for ensembles of particles, we analyze the statistics of the particle motion computed at various temperatures.
As expected, the mean displacement fluctuates around 0, while the root-mean squared displacement (RMSD) grows with time (see Fig.~\ref{fig:diffusion_rmsd}).
At very short time scales, $\tau \lesssim 0.1$, when the force experienced by the particle is roughly constant due to the slowly-changing potential profile, the RMSD exhibits a slope of $\sim 2$ on the log-log plot (shown as the dotted black line).
\begin{figure}
    \centering
    \includegraphics[width = \columnwidth]{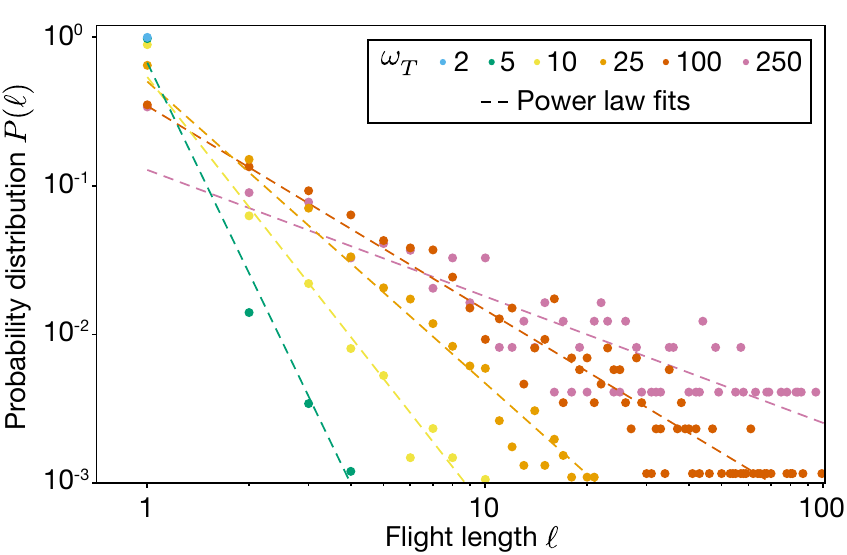}
    \caption{\emph{Probability distribution of flight lengths at various temperatures.} Using the 20 $\tau_0 = 100$ single-particle trajectories calculated above, the distribution of flight lengths $\ell$ for all temperatures (shown as data points of different colors) collapse onto lines of varying slopes on log-log axes. The dashed lines plot the best fit lines for each temperature and show the decrease in slope magnitude with temperature (from $-4.7$ at $\omega_T = 2$, to $-0.85$ at $\omega_T = 250$). At $\omega_T = 1$, particles remain trapped for the simulation duration and do not experience any flights.}
    \label{fig:flight_length}
\end{figure}

For longer times, at higher temperatures, the RMSD grows nearly linearly with time, indicating superdiffusive (nearly ballistic) motion.
On the log-log plot of Fig.~\ref{fig:diffusion_rmsd}, ballistic motion would exhibit a slope of $+1$ (shown as dashed lines), while Fickian diffusion would lead to a line with slope $+1/2$.
For the lowest temperatures, the particles are trapped in energy minima, resulting in an almost constant RMSD. 
For intermediate temperatures, the occasional long flights of individual particles lead to jumps in the RMSD.
For larger ensembles, these individual jumps would dominate the RMSD less, and we would expect to see a smoother curve.
The slope of the curve is likely to be temperature-dependent, approaching ballistic at high temperatures.

As we can see in both the individual trajectories of Fig.~\ref{fig:unfolded_traj} and the ensemble properties of Fig.~\ref{fig:diffusion_rmsd}, particle motion is qualitatively different at low and high temperatures.
This behavior is reminiscent of results described in Ref.~\citep{Deng2017}, where the authors reported different regimes of ionic motion.
At low temperatures, the mobile ions remain frozen at their energy minima.
Raising the temperature allows the ions to occasionally hop to neighboring minima.
Finally, at high temperatures, the mobile ion sublattice effectively melts, resulting in a ``superionic flow."

From our results, we observe that, at low temperatures, the particle behavior is dominated by flights of short lengths, while higher flight lengths become more frequent at higher temperatures.
In other words, another way to characterize the transport properties of the system \emph{at all temperatures} is by analyzing the distribution of flight lengths.
We define the flight length $\ell$ as the number of chain masses a particle passes between each pair of turning points (i.e. when the velocity $\dot{\sigma}$ changes sign).
Fickian diffusion has only step lengths $\ell=1$, which would lead to a delta-function distribution $P(\ell)$.
Using the same trajectories as above, we calculate the probability distribution of flight lengths, $P(\ell)$, for various temperatures (see Fig.~\ref{fig:flight_length}).
We see that the distributions, when plotted on log-log axes, collapse onto straight lines with slopes dependent on the temperature, indicating a temperature-dependent power-law decay.
The dashed lines in Fig.~\ref{fig:flight_length} are best-fit lines for each temperature.
The slope magnitude decreases with temperature from $-4.7$ to $-0.85$---longer flights are more prevalent as temperature increases.
This power-law distribution is characteristic of L\'evy flights, which have been extensively studied in other contexts~\cite{Chechkin2008}.

\section{Summary}
\label{sec:summary}
Using a simple model for mobile particles interacting with a lattice in 1D, we explored how the interplay between thermal lattice vibrations and particle-lattice interactions influences transport properties.
We showed that the statistical properties of dissipation are governed on the scale of the mobile particle passing a single lattice mass.
This single-pass loss follows a normal distribution with mean similar to the loss in the non-thermal case, and a variance that grows with the chain temperature.
Many of the characteristics of this dissipative motion can be calculated using a simple velocity random-walk model.
In the presence of a bias, the mobile particles show signatures of drift velocities at low enough temperatures.
For an ensemble of particles initially at rest, we found that they thermalized with the chain, and their resulting motion at sufficiently high temperatures was superdiffusive, showing characteristics of L\'evy flights.

Returning to the question of what makes a good ionic conductor, our calculations suggest that the temperature dependence of particle transport is non-trivial.
We found that when the ionic motion is primarily dissipative, the mean energy loss is relatively temperature-independent, and thermal vibrations primarily contribute stochastic noise.
At high temperatures, a combination of two factors---higher activation probability and higher speed after activation---results in mobile particles with L\'{e}vy flight character due to decreased drag at high speeds.
This behavior suggests that 1D ionic transport is improved at higher temperatures, with a faster increase than Arrhenius scaling arguments would predict.

Real systems have framework masses that can move in more than one direction---this motion has consequences both for the band structure of the lattice, and the interaction potentials that are possible to simulate.
In future work, we plan to incorporate these additional degrees of freedom into the equations of motion, and to employ Coulomb-like interactions.
Looking ahead to eventual extensions into higher dimensions, we also aim to improve the computational efficiency of our calculations.

\acknowledgments

H.M. is supported by the Ministry of Education, Singapore, under its Research Centre of Excellence award to the Institute for Functional Intelligent Materials (I-FIM, Project No. EDUNC-33-18-279-V12).
B.A.O. acknowledges support from the Ministry of Education and Yale-NUS College (through Start-up grant and Grant Nos. IG20-SG102 and IG20-SI101). 
A.R. acknowledges the National Research Foundation, Prime Minister Office, Singapore, under its Medium Sized Centre Programme and the support by Yale-NUS College (through Start-up Grant).
The computational work involved in this project was partially supported by NUS IT Research Computing Group.

We performed all our calculations using the {\scshape julia} programming language~\citep{Bezanson2017} and our code is available at https://github.com/rodin-physics/1d-ionic-chain.
Due to their size, the output files are not included in the repository.
Our plots are visualized using Makie.jl,~\citep{Danisch2021} employing a color scheme suitable for color-blind readers, developed in \citep{Wong2011}.

H.M., B.A.O., and A.R. conceptualized the work; H.M. and A.R. wrote the code which was used by them to run the simulations; H.M., B.A.O., and A.R. analyzed the results, prepared the graphics, and wrote the manuscript.

\appendix
\section{Effect of temperature on potential profile}
\label{sec:temperature_effect}
\begin{figure}
    \centering
    \includegraphics[width = \columnwidth]{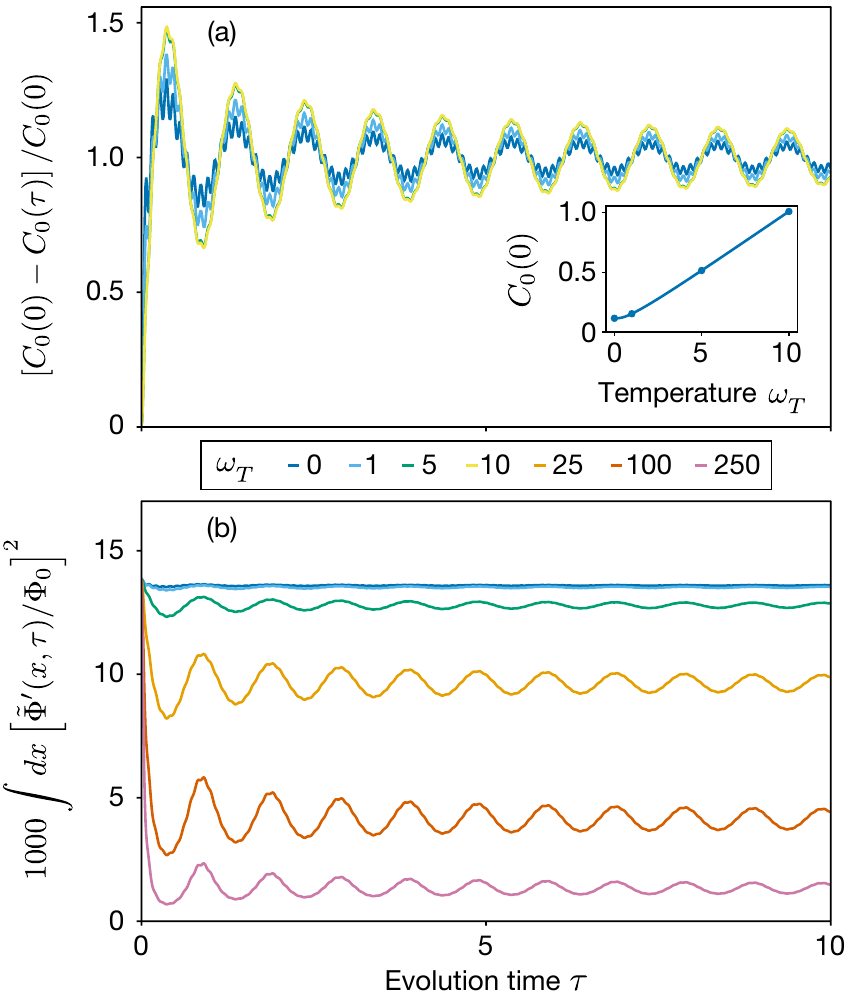}
    \caption{\emph{Effect of temperature on chain-particle interaction.} (a) variation of chain-particle interaction broadening parameter arising from thermal vibrations. The broadening factor is normalized with respect to $C_0(0)$ (shown in inset), which corresponds to the chain mass self-correlation at $\tau = 0$. The integral of the real-space broadened chain particle in plotted in (b), normalized with respect to $\Phi_0$ and showing how the chain-particle interaction varies with time. The parameters used here are $\alpha = 40$, $\Phi_0 = 2$, $\lambda = 4$ and $\omega_\mathrm{fast} = 10$.}
    \label{fig:Broadening_parameter}
\end{figure}

To understand the nonmonotonicity in low-velocity single-pass dissipation as temperature increases, we plot the prefactor for $\dot{\sigma}^{-1}$ from Eq.~\eqref{eqn:F_eval} in Fig.~\ref{fig:Broadening_parameter}(b).
Rewriting the expression in real space shows that the prefactor is proportional to the average of the square of the force applied by the particle onto the chain mass.
When calculating $\langle\Delta\rangle$, the curves in Fig.~\ref{fig:Broadening_parameter}(b) are multiplied by $e^{2\pi i\omega_j\tau''}$ and integrated over $\tau$ and the mode frequencies, giving the coupling amplitude to individual chain modes.
Whether the particle couples strongly to any of the modes, and hence dissipates more energy, is dependent on two factors: the magnitude of the force and the fluctuation amplitude of the force. 
At low temperatures, the force does not fluctuate a lot and hence does not couple strongly to any of the modes.
At very high temperatures, the strong broadening of the potential reduces the magnitude of the force, making the coupling weaker, also suppressing the energy transfer to the chain.
Finally, at intermediate temperatures, the average force oscillates without being too broadened, maximizing the dissipation.
                        
\section{Effect of exclusion zone on dissipation statistics}
\label{sec:exclusion_zone}

\begin{figure}
    \centering
    \includegraphics[width = \columnwidth]{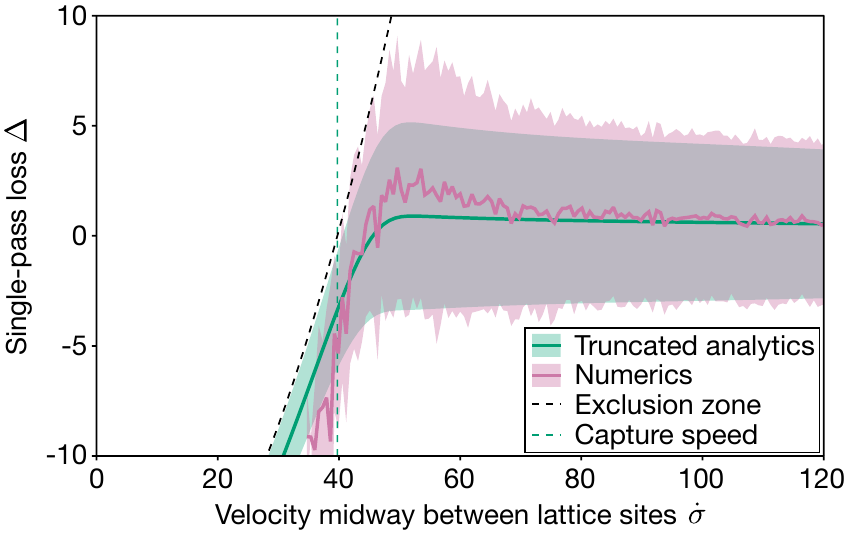}
    \caption{\emph{Energy loss distributions from analytical predictions and numerical full trajectory runs.} Distribution of energy losses after a single pass $\Delta$ predicted by Eqs.~\eqref{eqn:gaussian_mean_delta}, \eqref{eqn:gaussian_mean_var} and \eqref{eqn:truncated_gaussian}, shown as a green curve and band.
    The numerical data are identical to Fig.~\ref{fig:full_traj_loss}(c), namely, $~\sim250$ full trajectory calculations with memory parameter $\tau_0 = 100$, and are shown as a pink curve and band. 
    The vertical dashed green line is the capture speed of the system, and the dashed black curve shows the edge of the exclusion zone beyond which particles are unable to pass the chain mass.
    The numerical data exhibit a peak due to velocity fluctuations while passing each chain mass, but otherwise match the truncated analytic predictions well.}
    \label{fig:truncated_gaussian}
\end{figure}
In the thermal regime, the dissipation statistics along a multi-pass trajectory are the same as those of the single-pass case, as shown in Fig.~\ref{fig:full_traj_loss}. 
At speeds close to the capture speed, however, there is a sharp decrease in the numerical mean into negative values that is not predicted by Eqs.~\eqref{eqn:gaussian_mean_delta} and \eqref{eqn:gaussian_mean_var}. 
Since a value of $\Delta$ can be obtained only when the particle is able to pass a chain mass, the $\Delta$ normal distribution becomes truncated at its upper tail.
This deviation then reflects the statistics of particle trapping, rather than the dissipative properties of the system.
The truncation occurs at $\Delta_{\mathrm{ex}}(\dot{\sigma}) = \mu \dot{\sigma}^2 / (8 \pi^2) - \Phi_0$, which we will refer to as the edge of the exclusion zone.

We can analytically predict the effects of particle trapping by obtaining the mean and fluctuation of a Gaussian with one-sided truncation. For a Gaussian distribution $(\sigma \sqrt{2\pi})^{-1} \exp\left(\frac{(x - \mu)^2}{2\sigma^2}\right)$ truncated at $x = b$, the modified mean and variance are given by 
\begin{align}
    \tilde{\mu} &= \mu - \sigma \frac{\psi(\xi)}{\Psi(\xi)} \nonumber \\ 
    \tilde{\sigma}^2 &= \sigma^2 \left[1 - \xi \frac{\psi(\xi)}{\Psi(\xi)} - \left( \frac{\psi(\xi)}{\Psi(\xi)}\right)^2 \right] \, ,
    \label{eqn:truncated_gaussian}
\end{align}
where $\psi$ and $\Psi$ are the probability density function and cumulative distribution function, respectively, and $\xi = (b - \mu) / \sigma$~\cite{Johnson}. In our case, we substitute the mean and variance with $\langle \Delta \rangle$ and $\langle \Delta^2 \rangle$, and calculate the truncation value $b$ using $\Delta_\mathrm{ex}$. 

The results are plotted in Fig.~\ref{fig:truncated_gaussian}, with the green (pink) curve and band plotting the mean and fluctuation of the analytical prediction (numerical trajectory). 
For high speeds, the truncated results are identical to those of the full distribution (blue curve and band in Fig.~\ref{fig:full_traj_loss}), since a negligible portion of the distribution gets excluded. 
Close to the capture speed, the truncation causes the mean to decrease and the variance to be smaller.

The particle velocity $\dot\sigma$ varies during each pass: in the case of a repulsive potential, it slows down near each chain mass (see \cite{Mahalingam2023} for more discussion).
This variation leads the mean loss and fluctuation of the numerical calculations to be dominated by lower-velocity regions, and leads to the peak shown in Fig.~\ref{fig:truncated_gaussian}.
Other than this peak, the numerical and analytic curves match well, validating our hypothesis that the shift of $\langle \Delta \rangle$ towards negative values is caused by particle trapping.

\section{Random walk model}
\label{sec:random_walk}
\begin{figure}
    \centering
    \includegraphics[width = \columnwidth]{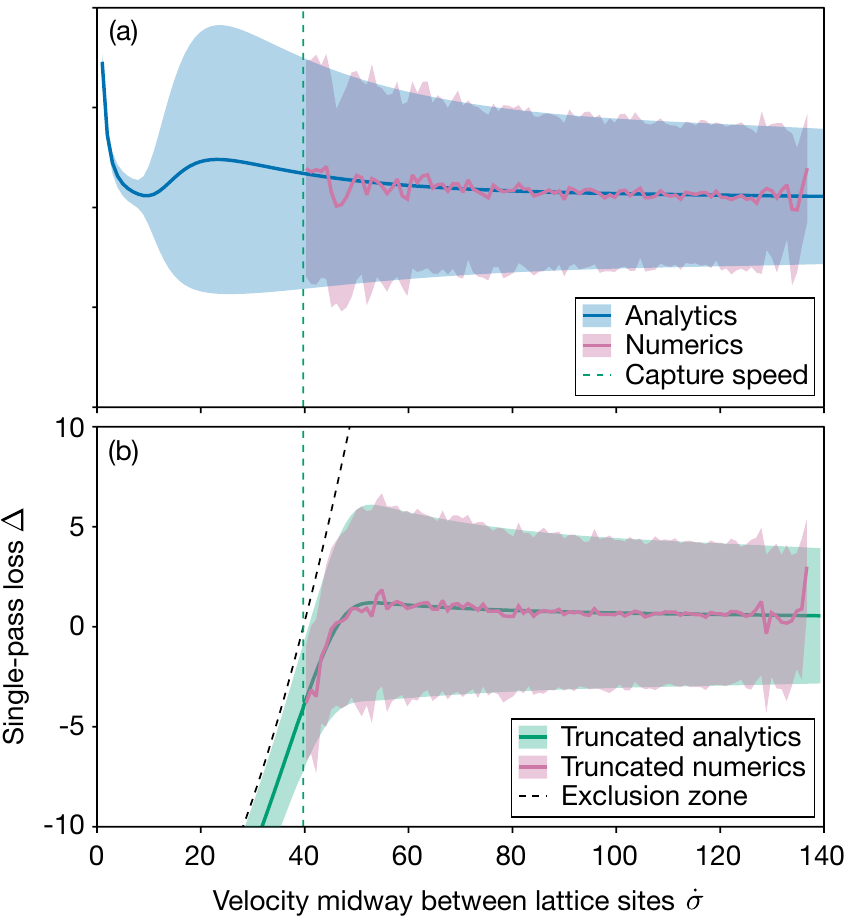}
    \caption{\emph{Energy losses $\Delta$ for full trajectories for a random walk model without and with exclusion zone.}  The energy losses after passing a chain mass extracted from a random walk model determined by the same system parameters as in Fig.~\ref{fig:full_traj_loss}. The pink curve and band plot the mean and variance of the numerical data, binned along speed, for $\sim270$ particles initialized at $\dot{\sigma} = 120$. The blue curve and band in (a) plots the analytic prediction given by Eqs.~\eqref{eqn:gaussian_mean_delta} and ~\eqref{eqn:gaussian_mean_var}, while the green curve and band in (b) plots the truncated analytic prediction given by Eq.~\eqref{eqn:truncated_gaussian}. The capture speed is plotted as the vertical green dashed line, while the exclusion zone edge is plotted as the black dashed curve.}
    \label{fig:random_walk}
\end{figure}
We have shown that the behavior of mobile particles is determined largely by the homogeneous motion of the chain masses rather than the deflection caused by the motion of the particle. 
In essence, a particle trajectory is given by a random walk in speed, with the statistics of energy loss after passing one chain mass given by Eqs.~\eqref{eqn:gaussian_mean_delta} and \eqref{eqn:gaussian_mean_var}, which describe a normal distribution. 
Specifically, for a particle initialized at speed $\dot{\sigma}_0$, the new speed after passing a single chain mass is 
\begin{equation}
\dot{\sigma} = \sqrt{\dot{\sigma}_0^2 - \frac{8\pi^2\Phi_0}{\mu} \textrm{N}(\langle \Delta \rangle, \sqrt{\langle \Delta^2 \rangle})} \, ,
\end{equation}
where we sample the energy loss from a normal distribution $\textrm{N}$. 
Employing this simple random-walk model for particles initialized at $\dot \sigma = 120$ (see Fig.~\ref{fig:random_walk}(a)), we see that the mean loss and fluctuations match the analytic predictions of Eqs.~\eqref{eqn:gaussian_mean_delta} and \eqref{eqn:gaussian_mean_var} well.

As discussed in Appendix~\ref{sec:exclusion_zone}, however, these statistics do not accurately describe particle motion near the capture velocity, as some of the particles will dissipate enough energy and become trapped between chain masses.
Following the approach of Appendix~\ref{sec:exclusion_zone}, we can take these captures into account by employing an exclusion zone in the velocity: once a particle's speed reaches this exclusion zone, we terminate its random-walk and exclude it from subsequent averaging.
In Figure~\ref{fig:random_walk}(b), we show the energy losses after passing a chain mass $\Delta$ for particles initialized at speed $\dot{\sigma}_0 = 120$ while employing this exclusion zone.
In Fig.~\ref{fig:random_walk}(a), the mean and variance computed with this random-walk model closely follow the analytic predictions of the single-pass model, while in Fig.~\ref{fig:random_walk}(b), the truncated random-walk dissipation closely matches the truncated analytic predictions from Appendix~\ref{sec:exclusion_zone}.

With the exception of the variation of the particle's speed during each pass, this simple model of a random walk in velocity with a truncated loss distribution reproduces all of the features of a full numerical trajectory.
Computationally, this simplified model is much less costly, and can be useful for efficiently predicting phenomena that are not dominated by trapping.


%

\end{document}